\documentclass[12pt,preprint]{aastex}

\received{}
\revised{}
\accepted{}
\journalid{}{}
\articleid{}{}
\paperid{}
\cpright{}{}

\ccc{}

\begin{document}

\title{Collective processes in relativistic plasma and their implications for
gamma-ray burst afterglows}

\author{Amir Sagiv\altaffilmark{1} and Eli Waxman\altaffilmark{1}}
\altaffiltext{1}{Department of Condensed Matter Physics, Weizmann
Institute, Rehovot 76100, Israel; amir,waxman@wicc.weizmann.ac.il}

\begin{abstract}

We consider the effects of collective plasma processes on synchrotron emission 
from highly relativistic electrons. We find, in agreement with Sazonov
(1970), that strong effects are possible also in the absence of a 
non-relativistic plasma component, due to the relativistic electrons 
(and protons) themselves. In contrast with Sazonov, who infers strong effects 
only in cases where the ratio of plasma frequency to cyclotron frequency 
is much larger than the square of the characteristic electron Lorentz factor, 
$\nu_p / \nu_B\gg \gamma_e^2$, we find strong effects also for 
$1\ll\nu_p / \nu_B\ll \gamma_e^2$. The modification of the spectrum is 
prominent at frequencies  
$\nu \leq \nu_{R^{\scriptstyle{*}}} \equiv \nu_p\; \min 
\left\{ \gamma_e, \sqrt{\nu_p / \nu_B}\, \right\}$, where 
$\nu_{R^{\scriptstyle{*}}}$ generalizes the \emph{Razin-Tsytovich} frequency, 
$\nu_R \equiv \gamma_e \nu_p$, to the regime $\nu_p / \nu_B \ll \gamma_e^2$.

Applying our results to $\gamma$-ray burst (GRB) plasmas, we predict a strong 
modification of the radio spectrum on minute time scale following the GRB,
at the onset of fireball interaction with its surrounding medium, in cases 
where the ratio of the energy carried by the relativistic electrons to the 
energy carried by the magnetic field exceeds $\sim 10^5$. Plausible 
electron distribution functions may lead to negative synchrotron reabsorption, 
i.e to coherent radio emission, which is characterized by a low degree of 
circular polarization. 
Detection of these effects would constrain the fraction of 
energy in the magnetic field, which is currently poorly determined by 
observations, and, moreover, would provide a novel handle on the properties 
of the environment into which the fireball expands.  
\end{abstract}


\section{Introduction}
\label{sec:Introduction}

According to the model now prevailing (see Piran 2000; M\'esz\'aros 2002 
for recent reviews), Gamma-Ray Bursts (GRB's) originate from the dissipation 
of the kinetic energy of a relativistically expanding fireball, caused by 
a cataclysmic collapse of a massive star or by a neutron star-neutron star or 
neutron star-black hole merger event, leading to the acceleration of a plasma
of electrons and protons to highly relativistic speed.
Part of the kinetic energy of the this expanding ``fireball'' is dissipated 
in ``internal'' collisions between different parts of the inhomogeneous ejecta,
resulting in shocks which accelerate particles via the Fermi process to 
ultra-relativistic energies. 
The non thermal radiation from accelerated electrons reproduces well 
the observed MeV $\gamma$-ray spectra. 
At a later stage of the expansion, the fireball
decelerates due to interaction with its surrounding medium. The relativistic
shock wave driven into the ambient medium continuously accelerates 
new electrons of the surrounding gas, producing a long-term synchrotron 
``afterglow'' emission. 

Although observations are in general agreement with model predictions
(see Kulkarni et al. 2000 for a recent review),
the model is incomplete, and there are several open issues, which are not 
properly understood. One is that of the burst progenitor. Several alternative 
models for the ``inner engine'' were suggested, such as NS-NS or NS-BH merger 
\cite{Paczynski1986, Goodman1986}, and the gravitational collapse of massive 
stars \cite{Woosley1993, Paczynski1998, MacFadyenWoosley1999}. 
Unfortunately, at present neither GRB nor afterglow observations provide
decisive evidence in support of a particular model. Nevertheless, the 
environment may be a clue to the progenitor. Thus, expansion into a 
relatively uniform inter-stellar medium (ISM) with a number density 
$n\sim1$~cm$^{-3}$ would be a natural consequence of a 
``merger'' scenario, whereas if the progenitor is a collapsing star, it is
natural to expect a much higher ambient density due to a wind ejected by 
the star at earlier stages of its evolution. The ``onset'' of fireball 
interaction with surrounding medium, i.e.
at the radius where fireball deceleration becomes significant, 
typically takes place on minute time scale (in the observer frame)
following the burst, at which stage the density of wind plasma is
$n\sim10^4$~cm$^{-3}$. At present, afterglow observations
typically begin several hours following the burst and do not allow to 
directly probe the onset of deceleration. Since on a day time scale
the fireball expands to a point where the wind density drops to
values close to that typical for the ISM, present observations do not
provide clear discriminants between the wind and ISM scenarios 
\cite{LivioWaxman2000}.

A second issue where basic understanding is still lacking is the physics 
of acceleration of electrons to high energies, and the build up of strong
magnetic fields by the GRB collisionless shock waves \cite{GruzinovWaxman1999}. 
The presence of high-energy electrons and strong magnetic fields is implied by 
observations, yet there is presently no theory based on first principles that 
satisfactorily explains electron coupling and magnetic field generation. 
The ignorance is usually parameterized by two dimensionless parameters, 
$\xi_e$ and $\xi_B$, which stand for the fractions of shock internal energy
density which are carried by the relativistic electrons and by the 
magnetic field, respectively. 
Efforts to constrain these parameters using afterglow 
observations are numerous, and $\xi_e$ is typically estimated to be close to 
its equipartition value, i.e. $\xi_e \sim1/3$. However, $\xi_B$ is not well 
constrained by observations, and its estimated values range from 
$\xi_B \sim 10^{-1}$ (e.g. Waxman 1997a; Wijers \& Galama 1998) to 
$\xi_B \sim 10^{-6}$ (e.g. Wijers \& Galama 1998; Chevalier \& Li 1999; 
Galama et al. 1999; Waxman \& Loeb 1999).

Observations strongly suggest that the radiation emitted during the afterglow 
is synchrotron radiation. We show here that the modification of the refractive 
index, by the relativistic electrons and protons, may strongly affect the 
emission at radio wavelengths during the onset of fireball deceleration, 
if $\xi_B\ll1$. Moreover, we find that, under plausible assumptions on the 
electron distribution function, the synchrotron reabsorption coefficient may 
become negative, thus leading to coherent emission from the fireball on minute 
time scale following the GRB. These collective plasma processes strongly 
affect the synchrotron spectrum at frequencies 
$\le\nu_{R^{\scriptstyle{*}}} = \nu_p \sqrt{\nu_p / \nu_B}$, 
where $\nu_p / \nu_B \sim \sqrt{\xi_e / \xi_B}$ is the ratio of plasma 
frequency and electron gyration frequency. 

As mentioned above, currently there is a few hours gap between GRB
and afterglow observations. However, the (operating) HETE-II satellite and 
the SWIFT satellite (planned to be launched at 2003)
may allow observations of early stages of fireball expansion, shortly after 
the GRB, thus providing data on the onset of deceleration. We show below that
observations of collective plasma effects in the emission from GRB afterglows 
will serve to constrain the value of $\xi_B$, as well as the parameters of 
the environment into which the fireball expands, 
thus providing a handle on both progenitor type and shock physics.

Collective plasma effects on synchrotron emission and reabsorption have 
been considered by many authors (e.g. Ginzburg \& Syrovatskii 1965; 
Zheleznyakov 1967; Yokun 1968; Crusius \& Schlickeiser 1988). Analyses of
effects due to a non-vacuum index of refraction are typically limited
to the case where the index of refraction is determined by the presence
of a non-relativistic plasma. These analyses do not apply to the GRB
plasma in which we are interested, where no ``cold''
non-relativistic plasma component is present. Sazonov
(1969, 1970, 1973) has studied 
the case of interest for us, where ``cold'' plasma is absent, and the effects 
are entirely due to the presence of the relativistic plasma.  
As we shall show, his results are too restrictive, as he infers 
negative reabsorption only in cases where $\nu_p / \nu_B \gg \gamma_e^2$, 
or in cases when the electron distribution function is very anisotropic. 
We find that these constraints might be eased, and that negative reabsorption 
is possible for isotropic distribution functions provided that 
$\nu_p \gg \nu_B$, thus allowing coherent emission for GRB 
plasma parameters.

In \S\ref{sec:VlasovTheory} we derive the dispersion relation and
the polarizations and refractive indices of normal modes
in a highly relativistic, weakly magnetized ($\nu_p/\nu_B\gg1$) plasma 
with isotropic electron and proton distribution functions. 
The frequency range discussed is $\nu_p<\nu\ll\gamma_e\nu_p\equiv\nu_R$, 
where the relativistic plasma may have a strong effect on synchrotron 
emission. In \S\ref{sec:RadiationProcesses} we show that the plasma strongly
affects synchrotron radiation at frequencies 
$\nu\le\nu_{R^{\scriptstyle{*}}} = \nu_p \sqrt{\nu_p / \nu_B}$, 
which generalizes the Razin-Tsytovich frequency $\nu_R$
to a regime where $\sqrt{\nu_p / \nu_B} \ll \gamma_e$ -- the regime relevant 
for us. We note, that although $\nu_R$ is commonly quoted as the frequency
below which plasma effects are strong, it is typically found in numerical
analyses of non-relativistic plasmas that, for 
$\nu_p / \nu_B\gg1$, synchrotron emission is
strongly modified only at $\nu\le\nu_{R^{\scriptstyle{*}}}$ 
(e.g. Crusius \& Schlickeiser 1988). 
We show in \S\ref{sec:RadiationProcesses}
that a more careful statement of the qualitative criterion, that leads to 
the conclusion that plasma effects are strong below $\nu_R$ 
(e.g Rybicki \& Lightman 1979 section 8.3), 
leads directly to the result that 
(for both relativistic and non-relativistic plasma) 
such effects are important only below $\nu_{R^{\scriptstyle{*}}}$.

Our results are applied to GRB plasmas in \S\ref{sec:GRBs}.
In \S\ref{sec:GRBPhenomenologyModel} we briefly describe GRB 
and afterglow phenomenology, give a short review of the fireball model,
and derive the plasma parameters during the onset of fireball deceleration, 
considering both expansion into a uniform-density ISM (\S\ref{sec:uniform}), 
and expansion into a wind (\S\ref{sec:wind}). 
Plasma effects are discussed in \S\ref{sec:FreqEstimates}. 
The implications of our results are discussed in \S\ref{sec:Discussion}.

We note, that coherent radio emission from GRB's has recently been discussed
by Usov \& Katz \cite{UsovKatz00}. The scenario considered by these authors 
for GRB production, the scattering of ambient medium electrons and protons by
a magnetically dominated wind \cite{Smolsky00}, 
is different, however, than the scenario we are 
considering, where observed radiation is produced (both in GRB and afterglow 
phases) by the dissipation through collisionless shocks of fireball kinetic
energy, leading to magnetic field amplification and to
particle acceleration. The processes we consider are, thus, different 
than those considered by Usov \& Katz (2000), and 
the predicted radio emission is 
very different. For example, while Usov \& Katz
find power-law spectrum with strong emission at $<1$~MHz for strong
magnetic field, we find strong emission only
for weak magnetic fields and over a narrow range of frequencies around
$\sim0.1$~GHz (which is more readily observable).

Finally , we note that the analysis presented here
is restricted to isotropic electron and proton distributions, 
and we show that strong plasma effects on synchrotron radiation occur 
if $\nu_p \gg \nu_B$. We note however, that anisotropy may lead to 
further interesting effects, the discussion of which is beyond the scope 
of this work.

\section{Waves in a weakly-magnetized relativistic plasma with 
isotropic electron and proton distribution functions}
\label{sec:VlasovTheory}

As explained in the next chapter, it is the deviation of the speed at 
which light propagates in plasma from $c$ which is responsible for the 
plasma effects we consider. This implies that we must first 
obtain expressions for the refractive indices $n_{1,2}(\omega)$ 
of the transversal electromagnetic modes in the plasma. In our derivation 
of the dispersion relation we restrict the discussion to plasmas
under the following conditions~:
(i) Electrons are highly relativistic; (ii) Rough energy
equipartition between protons and electrons; (iii) Isotropic particle
distribution functions; (iv) Weak magnetization, $\nu_p/\nu_B\gg1$.
Afterglow observations imply that (i) and (ii) hold for GRB
plasmas. 
(iii) restricts the discussion to plasma effects originating from the 
deviation of the refracting indices from their vacuum values. 
Our discussion is restricted to the weak magnetization case since,
as shown in the next chapter, strong effects 
of the plasma on synchrotron emission are obtained for this case only. 
Finally, we restrict the discussion to
the frequency range affected by collective effects,
$\nu_B\ll\nu_p<\nu\ll\gamma_e\nu_p$. 

The assumption $\nu_p/\nu_B\gg1$ allows a perturbative derivation (in 
$\nu_B/\nu_p$) of the dispersion relation.  
Moreover, in the frequency range $\nu_B\ll\nu_p<\nu\ll\gamma_e\nu_p$, 
the deviation of the refractive index 
$n=kc/\omega$, from 1 is much larger than $1/\gamma_e^2$. 
This simplifies the dispersion relations obtained below, 
thus allowing analytic estimates. Below we give order of magnitude estimates
of the plasma frequency and the difference between the refractive indices 
of the transversal electromagnetic waves in a relativistic plasma. 
Exact calculations are given in the appendix, 
and are used to verify the simple 
estimates.

As we show in \S\ref{sec:FieldFreePlasma}, the refractive indices of the 
transversal electromagnetic waves in a field-free plasma are degenerate, 
and satisfy approximately the relation 
\begin{eqnarray}\label{eq:PlasmaFrequency}
n^2(\nu) &=& 1-\nu_p^2/\nu^2,\nonumber\\
\nu_p &=& \frac{1}{2\pi}\left[\frac{4\pi n_e e^2}{\gamma_{e0}m_e} 
+\frac{4\pi n_p e^2}{\gamma_{p0}m_p}\right]^{1/2}\;,
\end{eqnarray}
where $\nu_p$ is the plasma frequency of a relativistic plasma. 
This result is exact for monoenergetic electron and proton distributions, 
where $\gamma_{e0}$ and $\gamma_{p0}$ are the associated Lorentz
factors of the two species, respectively. Detailed calculations show, 
however, that the above result is a good approximation to the plasma frequency 
also for more general energy distributions. 
For a power-law energy distribution implied by afterglow
observations, 
$n(\gamma)\propto\gamma^{-2}$ for $\gamma>\gamma_{min}$,
replacing $\gamma_0$ with $\gamma_{min}$ in equation (\ref{eq:PlasmaFrequency})
gives the plasma frequency with accuracy of 1\%.
Note, that equation (\ref{eq:PlasmaFrequency}) is a natural generalization to the 
relativistic regime of the familiar expression for the plasma frequency,
obtained by replacing the particle mass $m$ with $\gamma m$. 
Since we assume that the energy of the electrons and protons is approximately 
equipartitioned, the contributions of the two species to $\nu_p$ 
are comparable. 

Once an external magnetic field $\mathbf{B}_0$ is introduced, 
the plasma cannot be regarded as isotropic any longer, and the degeneracy 
of the two refractive indices is removed. For frequencies 
$\nu\ll\nu_B$ the effect of the magnetized plasma on the 
propagating radiation is small, and the electric field can be assumed to 
be approximately transversal to the direction of propagation, i.e. 
$\mathbf{E} \perp \mathbf{k}$. In the appendix (\S\ref{sec:WeaklyMagnetizedPlasma}) 
we derive the dispersion relation for the waves of transversal electric field.  
In order to obtain estimates of the refractive indices of the transversal 
EM modes, we solve the Vlasov equation of a relativistic plasma with isotropic
electron and proton distribution functions. 
The assumption of weak magnetization allows a perturbative expansion of the 
equation, where the perturbations in the particle distribution functions
are assumed to be linear in the external magnetic field, (i.e., the expansion 
parameter is $\nu_B/\nu_p$). This leads to a  significant simplification of the 
formalism. The dispersion relation for the components of the  transversal 
electric fields is given by~: 
\begin{equation}\label{eq:DispersionRel_temp}
\nonumber
\frac{k^2 c^2}{\omega^2} E_i = \epsilon_{ij} E_j \qquad\mbox{with} 
\quad i,j=1,2\;,
\end{equation}
where $\omega=2\pi\nu$ is the radian frequency of the wave, $k$ is 
the wave-number, and $\epsilon_{ij}$ is the $2\times2$ dielectric tensor. 
The refractive indices $n_{1,2}(\omega)=ck_{1,2}/\omega$ of the plasma are 
the square roots of the eigenvalues of this equation. 
Detailed numerical calculations showed that the values obtained for the components 
of the dielectric tensor and the resultant refractive indices are not sensitive to 
the exact shape of the electron and proton energy distribution functions.
Specifically, we showed that simple expressions obtained for monoenergetic 
distribution functions (which are the estimates used below) are good approximations 
of the more cumbersome expressions 
obtained for power law distributions. 

The deviation from degeneracy of the two refractive indices is the consequence 
of the non-diagonal terms of the dielectric tensor, which are proportional 
to $\cos\phi$, where $\phi$ is the angle between the wave-vector 
$\mathbf{k}$ and the external magnetic field $\mathbf{B}_0$ 
[eqs. (\ref{eq:epsilon11}) and (\ref{eq:epsilon12})~]. 
Hence the deviation is largest for radiation propagating along the magnetic field 
($\phi=0$), and vanishes when the direction of propagation is 
perpendicular to the magnetic field ($\phi=\pi/2$). 
Equations (\ref{eq:epsilon11_DeltaFunction_FieldFree}) and 
(\ref{eq:epsilon12_DeltaFunction}), describing the diagonal and off-diagonal
elements of the dielectric tensor,  
can be used to estimate the difference 
between the two refractive indices. For the frequencies in which we are 
interested $1-n$ is larger, or at least comparable to 
$1/\gamma_{e,p}^2$. This simplifies the algebra immensely. Since we 
assume that the widths of the electron and proton distribution functions 
are close, and that energy density in the protons and electrons is 
approximately equipartitioned, the contributions of the two species to 
$\Delta n=n_1-n_2$ are comparable. Hence, an order of magnitude estimate 
of $\Delta n$ is~:
\begin{equation}\label{eq:Delta_n}
\Delta n \sim \frac{\nu_p^2 \nu_B}{\nu^3}\cos\phi \ln 
\left(\frac{\nu_p^2}{4\nu^2}\right)
\end{equation}
up to a factor of order~1. 

The result $\Delta n_{\phi=\pi/2}=0$ is a consequence of the fact 
that our derivation keeps only terms which are linear in the magnetic field, 
i.e. $1^{\mathrm{st}}$ order terms in $\nu_B/\nu_p$. If this 
assumption is eased, and the refractive indices of the ordinary and 
extra-ordinary modes are used to estimate \,$\Delta n\,_{\phi=\pi/2}$, 
we obtain $\Delta n_{\phi=\pi/2}=\frac{1}{2} (\nu_B/\nu)^2 
\left[1-\nu_p^2/2\nu^2\right]$. As will be shown 
in the next chapter, the frequency at which the plasma has a significant 
effect on the synchrotron emission is $\nu_{R^{\scriptstyle{*}}}= 
\nu_p \sqrt{\nu_p/\nu_B}$. This implies that for plasma
parameters of interest to us 
($\nu_p /\nu_B\lesssim 10^3$) the logarithmic factor in 
equation (\ref{eq:Delta_n}) is of order 10. Hence we replace equation 
(\ref{eq:Delta_n}) with an expression which takes into account the finite 
$\Delta n$ at perpendicular propagation~:
\begin{equation}\label{eq:Delta_n_revised}
\Delta n(\phi) \simeq 
  \cases{\displaystyle\frac{\nu_p^2 \nu_B}{\nu^3} \cos\phi 
         \ln\left(\frac{\nu_p^2}{4\nu^2}\right) ,& 
         if $\displaystyle\left|\frac{\pi}{2}-\phi\right| > 
         \frac{1}{10} \frac{\nu_B\,\nu}{\nu_p^2}$ ;  \cr
         \displaystyle\frac{1}{2} \frac{\nu_B^2}{\nu^2} 
         \left[1-\frac{\nu_p^2}{2\nu^2}\right] ,& otherwise.\cr
         }
\end{equation} 

We solved the dispersion relation [eqs.  
(\ref{eq:epsilon11_DeltaFunction_FieldFree}) 
and (\ref{eq:epsilon12_DeltaFunction})] numerically, for two values of 
$\gamma_e$ (920 and $4.6\times10^4$), and for three 
values of $\nu_B/\nu_p$ ($10^{-1}, 10^{-2}$ and  
$10^{-3}$). These values of $\gamma_e$ and $\nu_B/\nu_p$
are representative of typical values for GRB afterglow plasmas (see 
\S\ref{sec:GRBs}). 
In these calculations we assumed a propagation angle $\phi = 
\pi/4$. The results agree with the estimate (\ref{eq:Delta_n_revised}) 
up to a factor of $\sim 2$. 
The numerical calculation also confirmed that for parameters in the ranges relevant
for GRB afterglows, the discrepancy between the refractive indices of the transversal 
modes is small in comparison to the deviation of either of them from unity, i.e.,
$|n_1-n_2| \ll (1-n_{1,2})$.  

When the radiation propagates parallel to the magnetic field, the squared 
refractive indices of the two normal modes may be described by the well-known 
expressions
\begin{eqnarray}\label{eq:nRnL_FluidPlasma}
\epsilon_R = 1 - \frac{\nu_{pe}^2}{\nu(\nu-\nu_{Be})} - 
\frac{\nu_{pp}^2}{\nu(\nu+\nu_{Bp})}
 \nonumber\\
\epsilon_L = 1 - \frac{\nu_{pe}^2}{\nu(\nu+\nu_{Be})} - 
\frac{\nu_{pp}^2}{\nu(\nu-\nu_{Bp})}\;,
\end{eqnarray}
where $\nu_{Be}, \nu_{Bp}$ are the gyration frequencies of electrons 
and protons in the magnetic field, and $\nu_{pe}, \nu_{pp}$ are the 
relativistic plasma frequencies of the two species, respectively.
It is instructive to compare our estimate in equation 
(\ref{eq:Delta_n_revised}) to the difference $n_R-n_L$ obtained from 
equation (\ref{eq:nRnL_FluidPlasma}). When exact equipartition is assumed, 
and both the protons and the electrons are mono-energetically distributed, 
$\nu_{pe}=\nu_{pp}=\nu_p$ and $\nu_{Be}=\nu_{Bp}=\nu_B$, whence 
$\epsilon_R=\epsilon_L=1 - 2\nu_p^2/(\nu^2-\nu_B^2)$. 
This degeneracy is removed if equipartition is not 
exactly satisfied, or if the widths of the distribution functions of the two 
species are different. Since both effects are expected to be important in the 
case of our interest, we will use equation (\ref{eq:Delta_n_revised}) as an 
estimate for $\Delta n$.

Our calculations imply that non-diagonal elements of the dielectric tensor 
are smaller than its diagonal elements by a factor of the order of $\nu_B/\nu\ll1$. 
The latter, however, are independent of the magnetic field, and thus are 
identical to the values we obtained  for a field-free plasma 
[see eq. (\ref{eq:epsilon11})]. 
As will be discussed in \S\ref{sec:RadiationProcesses}, the difference
between the refractive indices of the transversal modes is important
for determining the polarization of the modes and thus for the calculation
of synchrotron self-absorption. However, our estimate of the frequency
at which plasma effects on synchrotron emission become significant is not
sensitive to $\Delta n$. Therefore, as far as 
estimating the frequency $\nu_{R^{\scriptstyle{*}}}=\nu_p 
\sqrt{\nu_p/\nu_B}$ is concerned, we will use the approximation 
given in equation (\ref{eq:PlasmaFrequency}).

The detailed expressions we obtained (in the appendix) for the elements
of the $2\times2$ dielectric tensor $\epsilon_{ij}$, show that 
$\epsilon_{11}=\epsilon_{22}$ and $\epsilon_{12}=-\epsilon_{21}$ 
[see eqs. (\ref{eq:epsilon11_DeltaFunction_FieldFree}) and 
(\ref{eq:epsilon12_DeltaFunction})]. 
The assumption of transversality thus implies that the normal modes are 
always circularly polarized (clockwise and anti-clockwise). 
To check the consistency of our assumption of quasi-transversality, 
and to obtain the (elliptical) polarization of the two 
quasi-transversal modes, we calculated the eigenmodes of the full 
$3\times3$ dielectric tensor. Following the method outlined 
in \S\ref{sec:FieldFreePlasma} [and described in the paragraph preceding
eq. (\ref{eq:DispersionRel_temp})], it is straightforward (yet lengthy) 
to obtain expressions for the other elements of $\epsilon_{ij}$ 
with $i,j=1,2,3$, and \emph{without} assuming transversality. It is 
possible to show that these extra elements contribute an additional term 
to $\Delta n$, which is smaller than the value obtained from the 
transversal calculation by a factor of $\sim 0.1(\ln |\nu_B/ 
4\nu_p|)^{-4} \nu_p/\nu_B\sim10^{-2}$ at small and mildly 
oblique propagation angles, and becomes comparable to $\frac{1}{2}\, 
(\nu_B^2/\nu^2) \left[ 1-(\nu_p^2/2\nu^2)\right]$, as $|\pi/2-\phi|$ 
approaches $0.1\,\nu_B\nu/\nu_p^2$.
Numerical calculation of the plasma normal modes confirmed that for 
plasma parameters of interest to us ($\gamma_e=10^3-10^4, 
\nu_p/\nu_B=10^2-10^3$) the normal modes are indeed transverse 
to the direction of propagation, and are left- and right- circularly polarized.

\section{Effects on synchrotron emission}
\label{sec:RadiationProcesses}

The spectral characteristics of synchrotron emission by a single relativistic 
electron are determined by the beaming of the radiation emitted by the 
electron into a narrow cone about the instantaneous direction of the 
electron's motion. When the electron is in a vacuum, the opening angle of 
this cone depends on the electron's Lorentz factor and the frequency of the 
radiation. Three regimes should be considered: when the frequency of 
interest is higher than the characteristic synchrotron frequency 
$\nu_c$ the cone's opening angle is $\Delta\theta\simeq 
(\nu_c/\nu)^{1/2} /\gamma$, and when $\nu \sim \nu_c$, one obtains 
the familiar result $\Delta\theta\simeq 1 /\gamma$; If, however, the 
frequency of interest is much smaller than $\nu_c$, one obtains 
$\Delta\theta\simeq  (\nu_c/\nu)^{1/3} /\gamma$. Next we consider an 
electron which is embedded in a dielectric medium with a refractive index 
$n \neq 1$. Since the speed of light is now $c/n$, the opening angle of 
the cone is $\Delta\theta \simeq \left[(\nu_c/\nu)(1/\gamma^2) + 
1-n^2(\nu)\right]^{1/2}$ for $\nu>\nu_c$, $\Delta\theta \simeq 
\left[1/\gamma^2 + 1-n^2(\nu)\right]^{1/2}$ for frequencies  $\nu \sim 
\nu_c$, and $\Delta\theta \simeq \left[(\nu_c/\nu)^{2/3} (1/\gamma^2) 
+ 1-n^2(\nu)\right]^{1/2}$ for frequencies much smaller than $\nu_c$. 
When the dielectric medium is a plasma, we use $1-n^2=\nu_p^2/\nu^2$. 
A strong deviation from the vacuum behavior occurs when the second term 
$(\nu_p/\nu)^2$ is comparable to the first term in the square brackets. 
Using $\nu_c=\gamma^3\nu_B$, where $\nu_B$ is the gyration 
frequency of the electron, we obtain an expression for the frequency at which 
the plasma collective effects greatly influence the synchrotron emission~:
\begin{equation}\label{eq:RazinFrequency1}
\nu_{R^{\scriptstyle{*}}} \simeq \nu_p 
\min \left\{\gamma, \sqrt{\frac{\nu_p}{\nu_B}} \right\}\;.
\end{equation}
The \emph{Razin-Tsytovich} frequency $\nu_R=\nu_p\gamma$ is usually 
given as an estimate to the frequency below which the plasma strongly affects 
synchrotron radiation. Equation (\ref{eq:RazinFrequency1}) shows that this 
estimate is not generally valid. We are interested in the regime 
$\sqrt{\nu_p/\nu_B}\ll\gamma_e$, hence for plasma parameters 
prevailing in GRB afterglows the frequency at which the plasma greatly 
affects the synchrotron radiation is 
$\nu_{R^{\scriptstyle{*}}}=\nu_p\sqrt{\nu_p/\nu_B}$.
It is simple to show that the definition of the plasma frequency 
[eq. (\ref{eq:PlasmaFrequency})] implies that $\nu_p^2/\nu_B^2= 
\xi_e/2\ell\xi_B$, where $\ell=\ln ( \gamma_{e\,max} / 
\gamma_{e\,min} )$. Typically $\ell\sim4$. Thus
\begin{equation}
\label{eq:nu_R*}
\nu_{R^{\scriptstyle{*}}} = \nu_p \sqrt{\frac{\nu_p}{\nu_B}} = 
\nu_p\left( \frac{1}{2 \ell}
\frac{\xi_e}{\xi_B}\right)^{1/4} \;.
\end{equation}

An immediate consequence of equation (\ref{eq:nu_R*}) is that for the effect 
to be apparent we must have $\nu_p\gg\nu_B$ (or, equivalently, 
$\xi_e\gg2\ell\xi_B$). Sazonov (1970) also considered the 
effects of a relativistic plasma on synchrotron emission. However, he 
considered only a situation in which $\nu\sim\nu_c\ll\gamma \nu_p$, 
which led him to infer that negative synchrotron reabsorption is possible only 
for $\nu_p/\nu_B \gg \gamma_e^2$, or equivalently, if 
$\xi_e/\xi_B \gtrsim \gamma_e^4$. The result (\ref{eq:nu_R*}) 
shows that this condition is far too restrictive. However, as will be 
discussed in \S\ref{sec:FreqEstimates}, for the effects to be within 
the detection range of current radio telescopes, we must have 
$\xi_e/\xi_B \gtrsim 10^5$.

\subsection{The Razin suppression}
\label{sec:RazinEffect}

All the information regarding the response of the plasma to electromagnetic 
waves (e.g. modes propagating through the plasma, loss and gain, etc.)
is contained in the dielectric tensor. Specifically, the strong effect of 
plasma on synchrotron radiation is a manifestation of the deviation of 
the refractive index from 1, that is the non-trivial (i.e. ``non-vacuum'') 
structure of the dielectric tensor. The expression derived in the appendix 
for the dielectric tensor is not complete, i.e. it does not include
self-absorption, since we treat the magnetic field perturbatively.
Unfortunately, obtaining a complete expression for the dielectric tensor 
requires solving a kinetic equation for the plasma, a process which may 
be avoided by introducing the method of \emph{Einstein coefficients}.
The application of this method requires that over 
the length-scale of the wave phenomenon (i.e. the wavelength), the absorption
does not change the wave characteristics considerably. Indeed, since we do
not obtain the dielectric tensor explicitly, we do not know \emph{a priori} 
the EM modes which can propagate through the plasma. We must assume, however, 
that whatever these modes are, the modifications introduced to the dielectric 
tensor due to the presence of the plasma are dominated by the deviation of the 
refractive indices from unity, and not by the absorption, that is~:  
\begin{equation}\label{eq:ConsistencyCondition1}
\left| 1 - n^{(\mathit{l})}(\nu)\right| \gg \left| 
\alpha^{(\mathit{l})} (\nu) \lambda \right| \;,
\end{equation}
where $\lambda=c/\nu$ is the wavelength, and $l$ denotes the mode. 
When condition (\ref{eq:ConsistencyCondition1}) is not satisfied 
one must use a kinetic-equation approach, to self-consistently derive a dispersion 
relation which incorporates the self-absorption. 

Since the refractive index depends on the radiation mode under consideration, 
we now have to determine the modes relevant to the frequency range we are 
interested in. 
Motivated by the insight that in ``vacuum'' absorption is present, and 
the two transversal modes share the same refractive index (i.e., 1), whereas,
on the other hand, the result obtained above (see \S\ref{sec:VlasovTheory}), 
shows that when absorption is neglected one obtains a finite $\Delta n^{(l)}$ 
[see eq. (\ref{eq:Delta_n_revised})], we conclude that the 
question of determining the relevant modes is resolved by considering 
whether the following inequality holds~:
\begin{equation}\label{eq:IsotropyConsistencyCondition}
\left| n_1(\nu)-n_2(\nu)\right| \ll \left| 
\alpha_{\nu}^{(\mathit{1,2})}\lambda\right| \;,
\end{equation}
where $\alpha_{\nu}$ is the synchrotron self-absorption coefficient, 
and $\lambda$ the wavelength of radiation. 
When condition (\ref{eq:IsotropyConsistencyCondition}) is satisfied, 
the normal waves are linearly polarized along and 
perpendicular to the projection of the magnetic field on the plane of 
observation \cite{Ginzburg1989}. We denote these polarizations by 
$\perp$ and $\parallel$, respectively. The power emitted by a single 
electron is then given by 
\begin{mathletters}
\begin{eqnarray}\label{eq:RazinPowerPerpParallel}
P_{\perp,\parallel} (\nu, \gamma) &=& \frac{\sqrt{3} e^3 B\sin\chi}{2 m_e c^2} 
\left[1+\left(\frac{\gamma \nu_p}{\nu} \right)^2 \right]^{-1/2} 
\frac{\nu}{\widetilde\nu_c} \nonumber\\
&\times& \left[\int_{\nu/\widetilde\nu_c}^{\infty} K_{5/3}(z) dz  \pm 
K_{2/3}\left(\frac{\nu}{\widetilde\nu_c}\right)\right] \;, 
\end{eqnarray}
\begin{equation}
\label{eq:Plasma_nu_c_isotropic}
\widetilde\nu_c = \frac{3eB\sin\chi}{4\pi m_e c} \gamma^2 
\left[1+\left(\frac{\gamma \nu_p}{\nu}\right)^2 \right]^{-3/2}\;,
\end{equation}
\end{mathletters}
where $B$ is the strength of the magnetic field, and $\chi$ 
the pitch angle.

If, however, condition (\ref{eq:IsotropyConsistencyCondition}) is not satisfied, 
one cannot neglect the difference between the refractive indices of the normal modes. 
We are thus required to consider the two normal modes of the plasma, 
discussed previously in \S\ref{sec:VlasovTheory}. 
The normal modes are (quasi-) transversal, and are left- and right- circularly 
polarized. It can be shown \cite{Ginzburg1989}  that half the total power is 
``converted'' into each circularly-polarized normal wave (if we take into 
account terms up to order $1/\gamma$). In this case, the power emitted by 
a single electron is
\begin{mathletters}
\begin{equation}\label{eq:RazinPowerCircular}
P^{(1,2)}(\nu,\gamma) = \frac{\sqrt{3} e^3 B \sin\chi}{m_e c^2} 
\left[1+\gamma^2 (1-n^2_{1,2})\right]^{-1/2} \frac{\nu}{\widetilde\nu_c} 
\int_{\nu/\widetilde\nu_c}^{\infty} K_{5/3}(z) dz \;, 
\end{equation}
\begin{equation}\label{eq:Plasma_nu_c_unisotropic}
\widetilde\nu_c = \frac{3 e B \sin\chi}{4\pi m_e c} \gamma^2 
\left[1 + \gamma^2 (1-n^2_{1,2})\right]^{-3/2}\;.
\end{equation}
\end{mathletters}
The polarization of synchrotron emission in plasma will be further discussed in 
\S\ref{sec:SynchrotronPolarization}.

If the frequencies of interest are higher than $\nu_{R^{\scriptstyle{*}}}$, 
the emission does not differ greatly from the emission in vacuum. 
However, for frequencies $\nu \lesssim \nu_{R^{\scriptstyle{*}}}$ 
the emitted power is dramatically suppressed for both polarization regimes. 
This suppression is usually called the \emph{Razin} effect.

\subsection{Synchrotron self-absorption and the possibility of 
negative reabsorption}
\label{sec:NegativeReabsorption}

Our calculation of the synchrotron self-absorption coefficient $\alpha_\nu$ 
implements the Einstein relations between emission and absorption coefficients. 
These relations relate the emission and absorption coefficients of photons 
having a specific polarization state, or, to put it in terms of the previous 
chapter -- a specific mode propagating through the plasma, which we shall denote 
by $\mathit{l}$. Hence also the expression for the self-absorption coefficient 
is for a specific mode \cite{Ginzburg1989}~:
\begin{equation}\label{eq:SynchrotronSelfAbsorption}
\alpha_{\nu}^{\mathit(l)} = - \frac{c^2}{4\pi\nu^2} \int E^2 \frac{d}{d E} 
\left(\frac{n_e(E)}{E^2}\right) P^{(\mathit{l})} (\nu,E) dE \;,
\end{equation}
where $P^{(\mathit{l})} (\nu,E)$ is the power of $\mathit{l}-$polarized 
photons emitted by an electron with energy $E$.

Negative contributions to reabsorption come only from regions where the 
electron distribution function grows faster than $E^2$. Indeed, if the 
distribution function has regions which are ``steep'' enough, the 
self-absorption coefficient becomes negative at low frequencies 
($\nu\lesssim \nu_{R^{\scriptstyle{*}}}$). This  corresponds to stimulated 
emission from the plasma, and radiation is coherently amplified as it propagates. 
For this reason the effect is sometimes known as a \emph{maser} effect.

In order to estimate the ``amplitude'' of the negative self-absorption, we 
take the electron distribution to be mono-energetic. Substituting 
$n_e(E)=n_e \delta(E-E_0)$ in equation (\ref{eq:SynchrotronSelfAbsorption}), 
and using the fact that we are interested in frequencies $\nu\ll\gamma_e\nu_p$, 
it can be shown that 
\begin{equation}
\alpha_{\nu} \simeq 2\times10^{-2} (\nu_p \nu_B/c\sqrt{\gamma_e} \nu) 
\Phi(Z(\nu)) \;,
\end{equation}
where $\Phi(Z) = 2 Z \int_Z^{\infty} K_{5/3}(y) dy - 2Z K_{5/3}(Z)$, and 
$Z(\nu) = \frac{2}{3} \nu_p^3 /\nu_B \nu^2$. The function $\Phi(Z(\nu))$ 
has a global minimum at approximately $\nu_{R^{\scriptstyle{*}}}=\nu_p 
\sqrt{\nu_p/\nu_B}$, and its minimal value is $-0.24$. 
Substituting $\nu\sim\nu_{R^{\scriptstyle{*}}}$, we obtain an estimate of the 
minimal value of the synchrotron self-absorption~:
\begin{equation}\label{eq:min_alpha}
\min \alpha_{\nu} \approx - \frac{10^{-2}}{c\sqrt{\gamma_e}} \nu_B 
\sqrt{\frac{\nu_B}{\nu_p}}\;,
\end{equation}
up to a factor of order unity.

\subsection{Polarization of synchrotron radiation in plasma}
\label{sec:SynchrotronPolarization} 

We can now use the estimate of the absorption coefficient obtained above 
to check the self-consistency of our calculation. Namely, we shall show that in
the frequency range of interest, the deviation from unity of the refractive 
indices dominates the absorption, thus facilitating the validity of 
implementation of the Einstein coefficient method, as required by the consistency 
condition (\ref{eq:ConsistencyCondition1}).
In \S\ref{sec:FieldFreePlasma} we showed that the refractive indices may be 
approximated by $n\approx 1-\frac{1}{2}(\nu_p/\nu)^2$.  
Hence, for $\delta-$function distributions, equation (\ref{eq:ConsistencyCondition1}) 
may be re-written as $10^2 \sqrt{\gamma_e} \nu_p^{5/3}/\nu_B^{3/2} \gg \nu$. 
Since we are interested in frequencies of order of $\nu_{R^{\scriptstyle{*}}} 
\sim \nu_p \sqrt{\nu_p/\nu_B}$, this condition is casted to $10^2 \sqrt{\gamma_e} 
\nu_p/\nu_B \gg 1$, which is always satisfied for typical afterglow parameters 
($\gamma_e\sim 10^3, \nu_p/\nu_B \sim 10^3$). 

Next we determine the polarization of the synchrotron modes in the frequency range 
of interest for us. We use the estimate (\ref{eq:Delta_n_revised}) for $\Delta n$, 
and substitute equation (\ref{eq:min_alpha}) for the \emph{rhs} of the inequality 
(\ref{eq:IsotropyConsistencyCondition}) with negative reabsorption. After some  
algebra we find that condition (\ref{eq:IsotropyConsistencyCondition}) is satisfied 
only for radiation propagating almost perpendicular to the magnetic field, 
$|\pi/2-\phi|<0.1 (\nu_B/\nu_p)^{1/2}$, given that $\nu_B/\nu_p 
\ll 0.02 \gamma_e^{-1/2}$. 
For typical plasma parameters ($\gamma_e\sim 10^3-10^4, \nu_p/\nu_B \sim 10^3$) 
this  inequality is not satisfied, and we conclude that the plasma is not 
isotropic, hence the normal waves at frequencies $\nu \sim\nu_{R^{\scriptstyle{*}}}$ 
are circularly polarized.

\section{Application to GRBs}
\label{sec:GRBs}

This section is dedicated to the study of the application of the results 
obtained above to  the plasma conditions prevailing in GRB afterglows. 
We briefly review the fireball model, and obtain the physical parameters 
of the plasma during the onset of fireball deceleration, considering both 
expansion into a uniform density ISM (\S\ref{sec:uniform}), and into a wind 
(\S\ref{sec:wind}). These parameters enable order of magnitude estimates of 
relevant frequencies (\S\ref{sec:FrequencyEstimates_UniformISM} and 
\S\ref{sec:FrequencyEstimates_Wind}). We then make a short excursion, 
referring to the issue of the origin of negative reabsorption in relation 
to the shape of the electron distribution function at low energies 
(\S\ref{sec:DistributionFunction}), and conclude the section with a numerical 
calculation of the resultant spectrum, for various electron distribution 
functions (\S\ref{sec:Results}).

\subsection{Fireball plasma parameters during the onset of deceleration}
\label{sec:GRBPhenomenologyModel}

According to the fireball paradigm \cite{Paczynski1986, Goodman1986}, an 
energetic explosion ($E\sim10^{51}-10^{53}$~ergs) drives a relativistic 
blast wave into an ambient gas (the ``forward shock''). The expanding shell 
of accelerated ambient gas approaches gradually a self-similar behavior 
\cite{BlandfordMcKee1976}. During the short transition period before 
self-similarity is established, the interaction between the ejecta and the 
surrounding medium drives a relativistic shock into the ejecta 
(the ``reverse shock'') \cite{MeszarosRees1997}, and heats it. 
The transition to self-similarity occurs on a time scale comparable to the 
time it takes the reverse shock to cross the ejecta (e.g. Waxman \&
Draine 2000). 
The observed radiation is the consequence of synchrotron emission by 
relativistic shock-accelerated electrons, which gyrate in the magnetic 
fields generated by the shocks.

Recent data support GRB models where the outflow is a jet (rather than a sphere), 
with an opening angle $\theta_{jet}\sim0.1$
(Waxman, Kulkarni \& Frail 1998; Fruchter et al. 1999; Stanek et al. 1999;
Harrison et al. 1999; see Frail et al. 2001 for update analysis).
The dynamics and resulting light curves of such 
models differ from those of isotropic expansion models after the jet Lorentz 
factor decreases below $1/\theta_{jet}$. Typically this happens several hours 
after the main GRB. We, however, are interested in very early stages of the 
expansion (typically $\sim10$~s after the main GRB), and so the analysis we 
present below, although formulated in terms of an isotropic model, is valid 
also for a jetted scenario.

\subsubsection{Expansion into a uniform-density ISM}\label{sec:uniform}

Self similarity is established once the reverse shock crosses the ejecta. 
It has been shown \cite{WaxmanDraine2000} that at this time both the shocked 
ISM and the heated ejecta propagate with a Lorentz factor which is close to 
that given by the Blandford-Mckee self-similar solution~:
\begin{equation}
\Gamma^{(R)} \simeq \Gamma^{(F)} \simeq \left(\frac{17E}
{1024\pi n m_p c^5 T^3}\right)^{1/8} = 184 E_{52}^{1/8} n_0^{-1/8} 
T_{1}^{-3/8}\;,
\end{equation}
where $T=10 T_{1}$~s is the observed burst duration, which is typically of 
the order of $10$ seconds, $E=10^{52}E_{52}$~ergs is the explosion energy, 
and $n=1n_0$~cm$^{-3}$ is the density of the ambient gas. 
Accordingly, the electron number density behind the forward shock is 
\begin{equation}\label{eq:n_F}
n_e^{\prime (F)} = 4\Gamma^{(F)} n = 735 E_{52}^{1/8} T_{1}^{-3/8}
n_0^{7/8} \textrm{ cm}^{-3}\;,
\end{equation}
(where the prime denotes that this is a quantity measured in the frame 
comoving with the plasma). 

Let $\gamma_p^{(F)}$ and $\gamma_p^{(R)}$ be the Lorentz factors associated 
with the thermal motion of the protons accelerated by the forward and reverse 
shocks, respectively. 
Then it can be shown \cite{WaxmanDraine2000} that $\gamma_p^{(F)}\simeq 
\gamma_p^{(R)}\Gamma^2/\Gamma_i$, where $\Gamma_i\sim300$ is the Lorentz 
factor of the ejecta prior to its deceleration by the ambient gas 
(and after the production of the main GRB). 
This result holds for both relativistic and non-relativistic reverse shocks. 
Since the plasmas behind the forward and reverse shocks are separated by a 
contact discontinuity, the energy densities in the two plasmas are similar. 
Consequently we obtain
\begin{equation}
n_e^{\prime (R)} \simeq (\Gamma^2/\Gamma_i) n_e^{\prime (F)} = 
8.27\times10^4  E_{52}^{3/8} T_{1}^{-9/8} n_0^{5/8} \Gamma_{i\,2.5}^{-1} 
\textrm{ cm}^{-3}\;.
\end{equation} 

Lacking a fundamental theory of Fermi acceleration and formation of magnetic 
fields by shocks, it is customary to parameterize the fractions of the energy 
carried by the magnetic field and the electrons by two dimensionless parameters, 
$\xi_B$ and $\xi_e$, respectively. 
We assume that the values of these parameters are similar in the plasmas 
accelerated by the forward and reverse shocks, as we expect them to be 
associated with ``micro - physics'' processes in the plasma. 
Since energy densities behind both shocks are similar, we then have
\begin{equation}
B^{(R)} \simeq B^{(F)} = 0.07 \left(\frac{\xi_B}{10^{-6}}\right)^{1/2}
 E_{52}^{1/8} T_{1}^{-3/8} n_0^{3/8} \textrm{ G}\;,
\end{equation}
where we use the normalization $(\xi_B/10^{-6})$ since we expect strong 
collective plasma effects for small values of $\xi_B$ 
(see \S\ref{sec:Introduction}).

Relativistic shock waves are assumed to accelerate protons and electrons to 
high energies, giving power-law distribution functions~:
\begin{equation}\label{eq:PowerLawDistribution}
n_e^{\prime}(\gamma) = K \gamma^{-p}, \;\textrm{for}\; 
\gamma_{e\,min}<\gamma<\gamma_{e\,max}\;.
\end{equation}
Spectral indices $p\gtrsim2$ were observed in cosmic-rays spectrum and 
supernovae radio emission, observations which were later explained by 
theoretical work \cite{BlandfordEichler1987}.
Numeric and analytic calculations of particle acceleration via the first-order
Fermi mechanism in relativistic shocks yield spectral indices $p \approx 2.2$ 
for highly relativistic shocks \cite{BednarzOstrowski1998}. This result is in 
agreement with the value of $p$ inferred from GRB afterglow observations 
(Waxman 1997a, 1997b).
The maximum Lorentz factor $\gamma_{e\,max}$ is determined by requiring that 
the most energetic electrons lose energy through synchrotron emission slower than 
they gain energy due to acceleration by the shock. 
Normalization then implies
\begin{equation}\label{eq:gamma_e_min_Forward}
\gamma_{e\,min}^{\phantom{e\,}(F)} = \frac{1}{\ell} \xi_e 
\frac{m_p}{m_e} \Gamma = 4.22\times10^4 \ell_4^{-1} \left(\frac{\xi_e}{0.5}\right)
E_{52}^{1/8} T_{1}^{-3/8} n_0^{1/8}\;,
\end{equation}
where $\ell=\ln (\gamma_{e\,max}/\gamma_{e\,min})$. Typically $\ell\sim4$. 
Following the discussion leading to equation (\ref{eq:n_F}), we have 
$\gamma_{e\,min}^{\phantom{e\,}(R)} \simeq \gamma_{e\,min}^{\phantom{e\,}(F)} 
\Gamma_i/\Gamma^2$.

\subsubsection{Expansion into a wind}\label{sec:wind}

We examine here the simplest model of a ``wind'', in which the star loses mass 
at a constant rate $\dot{M}$ during an epoch prior to the explosion. 
Material is ejected radially at a constant speed $v$, hence  producing a 
non-homogeneous ambient gas with $n\propto r^{-2}$. 
Using typical estimates for the mass-loss rate 
$\dot{M}=10^{-5} \textrm{M}_{\odot}/\textrm{year}$ 
and for the wind velocity $v=10^3$~km~s$^{-1}$ \cite{ChevalierLi1999}, 
one obtains $n=\dot{M}/4\pi m_p v r^{-2} \simeq 3\times10^{35} 
\left(\dot{M}_{-5}/v_3\right) r^{-2}$~cm$^{-3}$; here we used the notations 
$\dot{M}_{-5} = \dot{M}/(10^5 \textrm{M}_{\odot}/\textrm{year})$, and 
$v_3= v/(10^3$~km~s$^{-1})$. 
Once the dynamics reaches the self-similar regime, the Lorentz factor and the 
radius are related through $E=(16\pi/9)m_p c^2 r^3 n(r)\Gamma_{B-M}^2 = 
(4\dot{M} c^2/9 v) \Gamma_{B-M}^2 r$. 
The physical parameters of the plasmas heated by the forward and reverse shocks 
are obtained following the reasoning employed in \S\ref{sec:uniform}, and so, 
without further ado: 
\begin{eqnarray}
n_e^{\prime (F)} &\simeq& 1.09\times10^7 E_{52}^{-3/4} (\dot{M}_{-5}/v_3)^{7/4}
T_{1}^{-5/4} \textrm{ cm}^{-3},\\
n_e^{\prime (R)} &\simeq& 6.56\times10^7 E_{52}^{-1/4} (\dot{M}_{-5}/v_3)^{5/4}
T_{1}^{-7/4} \Gamma_{i\,2.5}^{-1} \textrm{ cm}^{-3},\\
B^{(R)} &\simeq& B^{(F)} \simeq 4.18 \left( \frac{\xi_B}{10^{-6}} \right)^{1/2}
E_{52}^{-1/4} (\dot{M}_{-5}/v_3)^{3/4} T_{1}^{-3/4} \textrm{ G}, \\
\gamma_{e\,min}^{\phantom{e\,}(F)} &\simeq& 9.78\times10^3 \ell_4^{-1} 
\left( \frac{\xi_e}{0.5} \right) E_{52}^{1/4} (\dot{M}_{-5}/v_3)^{-1/4} 
T_{1}^{-1/4}\;,\\
\gamma_{e\,min}^{\phantom{e\,}(R)} &\simeq& 1.61\times10^3 \ell_4^{-1} 
\left( \frac{\xi_e}{0.5} \right) E_{52}^{-1/4} (\dot{M}_{-5}/v_3)^{1/4} 
T_{1}^{1/4} \Gamma_{i\,2.5}\;.
\end{eqnarray}

\subsection{Razin cutoff and synchrotron maser}
\label{sec:FreqEstimates}

We shall now use the lessons of the previous three chapters to estimate  
$\nu_p, \nu_B, \nu_c$ (the synchrotron characteristic frequency) and 
$\nu_{R^{\scriptstyle{*}}}$ for the four cases under study, i.e. the ejecta 
heated by the reverse shock, and the ambient medium heated by the forward shock, 
in the two fireball expansion scenarios.

\subsubsection{Expansion into a uniform-density ISM}
\label{sec:FrequencyEstimates_UniformISM}

The typical Lorentz factor of the ISM accelerated by the forward shock is very 
high, $\gamma_{e\,min}^{\phantom{e\,}(F)} \simeq 4.2\times10^4$, 
and the number density is $n_e^{\prime (F)} \simeq 735$~cm$^{-3}$. 
Under the assumption of a small $\xi_B$, the magnetic field is 
$B\simeq 0.07$~G. 
The values just stated correspond to the epoch of transition to self-similar 
behavior. Using the results of the last section, and boosting to the observer 
frame (multiplying by $\Gamma$), we obtain~:
\begin{eqnarray}\label{eq:UniformISM_ForwardShock}
\nu_p^{(F)} &\simeq& 2.2\times10^5 \ell_4^{1/2} \left(\frac{\xi_e}{0.5}\right)
^{-1/2} E_{52}^{1/8} T_{1}^{-3/8} n_0^{3/8} \textrm{ Hz} \nonumber\\
\nu_B &\simeq& 870 \ell_4 \left(\frac{\xi_e}{0.5}\right)^{-1} 
\left(\frac{\xi_B}{10^{-6}}\right)^{1/2} E_{52}^{1/8} T_{1}^{-3/8} 
n_0^{-1/8} \textrm{ Hz}\nonumber\\
\nu_c^{(F)} &\simeq& 9.8\times10^{16} \ell_4^{-2} \left(\frac{\xi_e}{0.5}
\right)^{2} \left(\frac{\xi_B}{10^{-6}}\right)^{1/2} E_{52}^{1/2} T_{1}^{-3/2} 
\textrm{ Hz}\nonumber\\ 
\nu_{R^{\scriptstyle{*}}}^{(F)} &\simeq& 3.5\times10^6 \ell_4^{1/4} 
\left(\frac{\xi_e}{0.5}\right)^{-1/4} \left(\frac{\xi_B}{10^{-6}}\right)^{-1/4} 
E_{52}^{1/8} T_{1}^{-3/8} n_0^{3/8} \textrm{ Hz} \;.
\end{eqnarray}
Since $\gamma_{e\,min}^{(F)}, B$ and $ n_e^{\prime (F)}$ are proportional to 
$\Gamma\propto r^{-3/2}$, we find that (in the observer's frame) 
$\nu_p, \nu_B$ and $ \nu_{R^{\scriptstyle{*}}} \propto r^{-3/2}$, 
whereas $\nu_c \propto r^{-6}$.

The ejecta heated by the reverse shock is characterized by a lower Lorentz 
factor $\gamma_{e\,min}^{\phantom{e\,}(R)} \simeq 375$, but the number density 
is much higher than behind the forward shock, $n_e^{\prime (R)} \simeq 
8.3\times10^4$~cm$^{-3}$. In the observer's frame~:
\begin{eqnarray}\label{eq:UniformISM_ReverseShock}
\nu_p^{(R)} &\simeq& 2.5\times10^7 \ell_4^{1/2} 
\left(\frac{\xi_e}{0.5}\right)^{-1/2} E_{52}^{3/8} T_{1}^{-9/8} n_0^{1/8} 
\Gamma_{i\,2.5}^{-1} \textrm{ Hz} \nonumber\\
\nu_B^{(R)} &\simeq& 9.8\times10^4 \ell_4 \left(\frac{\xi_e}{0.5}\right)^{-1} 
\left(\frac{\xi_B}{10^{-6}}\right)^{1/2} E_{52}^{3/8} T_{1}^{-9/8} n_0^{1/8}  
\Gamma_{i\,2.5}^{-1} \textrm{ Hz} \nonumber\\ 
\nu_c^{(R)} &\simeq& 7.7\times10^{12} \ell_4^{-2} 
\left(\frac{\xi_e}{0.5}\right)^{2} \left(\frac{\xi_B}{10^{-6}}\right)^{1/2} 
n_0^{1/2} \Gamma_{i\,2.5}^2  \textrm{ Hz} \nonumber\\   
\nu_{R^{\scriptstyle{*}}}^{(R)} &\simeq& 3.9\times10^8 \ell_4^{1/4} 
\left(\frac{\xi_e}{0.5}\right)^{-1/4} \left(\frac{\xi_B}{10^{-6}}\right)^{-1/4} 
E_{52}^{3/8} T_{1}^{-9/8} n_0^{1/8} \Gamma_{i\,2.5}^{-1} \textrm{ Hz} \;. 
\end{eqnarray}

\subsubsection{Expansion into a wind}\label{sec:FrequencyEstimates_Wind}

While the reverse shock crosses the ejecta, the heated wind particles are 
accelerated by the forward shock to a typical Lorentz factor 
$\gamma_{e\,min}^{\phantom{e\,}(F)} \simeq 9.8\times10^3$. 
At that time the number density is 
$n_e^{\prime (F)} \simeq 1.1\times10^7$~cm$^{-3}$, 
and the magnetic field is $B\simeq4.2$~G. Then~:
\begin{eqnarray}\label{eq:Wind_ForwardShock}
\nu_p^{(F)} &\simeq& 1.3\times10^7 \ell_4^{1/2} 
\left(\frac{\xi_e}{0.5}\right)^{-1/2} (\dot{M}_{-5}/v_3)^{3/4} E_{52}^{-1/4} 
T_{1}^{-3/4} \textrm{ Hz} \nonumber\\
\nu_B^{(F)} &\simeq& 5.1\times10^4 \ell_4 \left(\frac{\xi_e}{0.5}\right)^{-1} 
\left(\frac{\xi_B}{10^{-6}}\right)^{1/2} (\dot{M}_{-5}/v_3)^{3/4} 
E_{52}^{-1/4} T_{1}^{-3/4} \textrm{ Hz} \nonumber\\ 
\nu_c^{(F)} &\simeq& 7.1\times10^{16} \ell_4^{-2} 
\left(\frac{\xi_e}{0.5}\right)^{2} \left(\frac{\xi_B}{10^{-6}}\right)^{1/2} 
E_{52}^{1/2} T_{1}^{-3/2} \textrm{ Hz} \nonumber\\  
\nu_{R^{\scriptstyle{*}}}^{(F)} &\simeq& 2.0\times10^8 \ell_4^{1/4} 
\left(\frac{\xi_e}{0.5}\right)^{-1/4} \left(\frac{\xi_B}{10^{-6}}\right)^{-1/4} 
(\dot{M}_{-5}/v_3)^{3/4} E_{52}^{-1/4} T_{1}^{-3/4} \textrm{ Hz} \;.
\end{eqnarray}
Since during the self-similar stage $n_e^\prime$ scales as $\Gamma^5$, 
$B\propto\Gamma^3$ and $\gamma_{e\,min}\propto\Gamma$, and 
$\Gamma\propto r^{-1/2}$, we obtain the following scaling relations for 
later times~: 
$\nu_p, \nu_B$ and $\nu_{R^{\scriptstyle{*}}} \propto r^{-3/2}$, 
$\nu_c \propto r^{-3}$. 
However, for earlier times the forward shock expands at a uniform Lorentz 
factor, and so $\gamma_{e\,min}\propto\Gamma=$~const.,  
$n_e^\prime\propto\Gamma n\propto r^{-2}$, and 
$B\propto\sqrt{u^\prime} \propto \sqrt{\Gamma^2 n} \propto r^{-1}$. 
Hence we find that $\nu_p, \nu_B, \nu_c$ and 
$\nu_{R^{\scriptstyle{*}}}$ all scale as $r^{-1}$. 

And finally, the electrons in the ejecta heated by the reverse shock when 
the fireball expands into a wind have a Lorentz factor of 
$\gamma_{e\,min}^{\phantom{e\,}(R)} \simeq 1.6\times10^3$, 
and number density $n_e^{\prime (R)} \simeq 6.6\times10^7$~cm$^{-3}$. Hence~:
\begin{eqnarray}\label{eq:Wind_ReverseShock}
\nu_p^{(R)} &\simeq& 7.7\times10^7 \ell_4^{1/2} 
\left(\frac{\xi_e}{0.5}\right)^{-1/2} (\dot{M}_{-5}/v_3)^{1/4} 
E_{52}^{1/4} T_{1}^{-5/4} \Gamma_{i\,2.5}^{-1} \textrm{ Hz} \nonumber\\
\nu_B^{(R)} &\simeq& 3.1\times10^5 \ell_4 \left(\frac{\xi_e}{0.5}\right)^{-1} 
\left(\frac{\xi_B}{10^{-6}}\right)^{1/2} (\dot{M}_{-5}/v_3)^{1/4} 
E_{52}^{1/4} T_{1}^{-5/4} \Gamma_{i\,2.5}^{-1} \textrm{ Hz} \nonumber\\ 
\nu_c^{(R)} &\simeq& 2.0\times10^{15} \ell_4^{-2} 
\left(\frac{\xi_e}{0.5}\right)^{2} \left(\frac{\xi_B}{10^{-6}}\right)^{1/2} 
(\dot{M}_{-5}/v_3) E_{52}^{-1/2} T_{1}^{-1/2} \Gamma_{i\,2.5}^2 
\textrm{ Hz} \nonumber\\  
\nu_{R^{\scriptstyle{*}}}^{(R)} &\simeq& 1.2\times10^9 \ell_4^{1/4} 
\left(\frac{\xi_e}{0.5}\right)^{-1/4} \left(\frac{\xi_B}{10^{-6}}\right)^{-1/4} 
(\dot{M}_{-5}/v_3)^{1/4} E_{52}^{1/4} T_{1}^{-5/4} \Gamma_{i\,2.5}^{-1} 
\textrm{ Hz} \;.
\end{eqnarray}

\subsubsection{The electron distribution function at low energies}
\label{sec:DistributionFunction}

As stated in \S\ref{sec:NegativeReabsorption}, a necessary condition for 
negative reabsorption is a region which grows faster than $E^2$ in the electron 
distribution function. We give here a plausible mechanism which may be 
responsible for the existence of such a region. Relativistic shock waves are 
thought to accelerate electrons in such a way that the electron distribution 
function has a power-law tail extending to high energies. 
However, there is no satisfactory theory which predicts the shape of the 
distribution function at low energies. We thus assume that at low energies the 
distribution function has the simplest possible behavior, and electrons 
distribute according to the volume they occupy in phase-space, i.e. 
$n_e(E) \propto E^2$. By emitting synchrotron radiation, the electrons gradually 
lose their energy, and accumulate at lower energies, thus leading to a distribution 
function which grows faster than $E^2$ at low energies. 
The excess of electrons above the $E^2$ power law is sensitively dependent on 
the details of the distribution function injected by the shock. 
Consider, for example, an injected distribution function composed of two \emph{pure} 
power laws [i.e., $n_e(E)\propto E^2$ if $E\leq E_0$, $n_e(E)\propto 
E^{-p}$ if $E>E_0$]. This distribution function has a discontinuous derivative at its 
peak. Synchrotron cooling thus results in an excess of ``cooled'' electrons just 
below $E_0$, leading to a strong maser effect. This issue will be treated further 
in \S\ref{sec:Results}.
 
It must be borne in mind, though, that electrons which are ``injected'' 
to the hot plasma by the shock at later times have less time to cool. 
This leads, in principal, to a significant 
complication in the distribution function of the shock-accelerated electrons, 
since the further away from the shock front we look, the ``colder'' the 
distribution is, and in general, we cannot consider the distribution function 
of the shocked plasma to be homogeneous. In particular, in regions closest to 
the shock front, where the electrons had very little time to cool, compared 
to the dynamical time, the reabsorption is positive, and thus may obscure 
coherent emission from regions further away from the shock front. 
Nevertheless, we have shown numerically that these regions contribute an 
optical depth of order of few, at most, for typical afterglow parameters; 
hence the effect of inhomogeneity does not change qualitatively the 
phenomenon of coherent emission by the relativistic plasma. 
Furthermore, if one assumes some turbulent mechanism which acts on the 
dynamical time scale of the system, and ``mixes'' electron populations 
which were injected by the shock at different times, one may disregard 
that complication, and treat the ``averaged distribution'' function. 

It is interesting to notice that a major effect exists even when the 
low-energy electrons are distributed as $n_e(E)\propto E^2$~! 
Indeed, there is no negative reabsorption in this case, since 
$\alpha_{\nu}<0$ requires a region steeper than $E^2$ in the distribution 
function. Nevertheless, the emissivity at $\nu\sim\nu_{R^{\scriptstyle{*}}}$ 
is dominated by the low-energy electrons (since $j_{\nu}$ of electrons with 
$\gamma_e\geq\gamma_{e\,min}$ is Razin-suppressed at higher frequencies). 
On the other hand, the absorption coefficient, albeit always positive, is 
dominated by the high energy electrons, and so starts decreasing at frequencies 
higher than $\nu_{R^{\scriptstyle{*}}}$. 
Altogether, we obtain a high peak in the intensity for 
$\tau_{\nu_{R^{\scriptstyle{*}}}} \gg 1$, and a strong suppression when 
$\tau_{\nu_{R^{\scriptstyle{*}}}} \ll 1$, 
where $\tau_{\nu_{R^{\scriptstyle{*}}}}$ is the optical depth 
at~$\nu_{R^{\scriptstyle{*}}}$.

\subsubsection{Results of detailed calculations}\label{sec:Results}

We calculated the emitted intensity for a plasma with 
$\gamma_{e\,min}=10^3-10^4, \nu_p/\nu_B=10^2-10^3$. 
Following the discussion in \S\ref{sec:SynchrotronPolarization}, 
the normal waves propagating in the plasma are circularly polarized. 
The intensity of the emitted radiation is given by 
\begin{equation}
I_{\nu(1,2)} = j_{\nu(1,2)} \Delta^\prime 
\frac{1 - e^{-\tau_{\nu(1,2)}}}{\tau_{\nu(1,2)}} \;,
\end{equation}
where $j_{\nu}$ is the specific emissivity, and $\tau_{\nu}$ the optical depth.
The width (as measured in the comoving frame) of the emitting medium along the 
line of sight $\Delta^\prime$ is typically $\sim10^{13}$~cm in both expansion 
scenarios. 

\begin{figure}
\plottwo{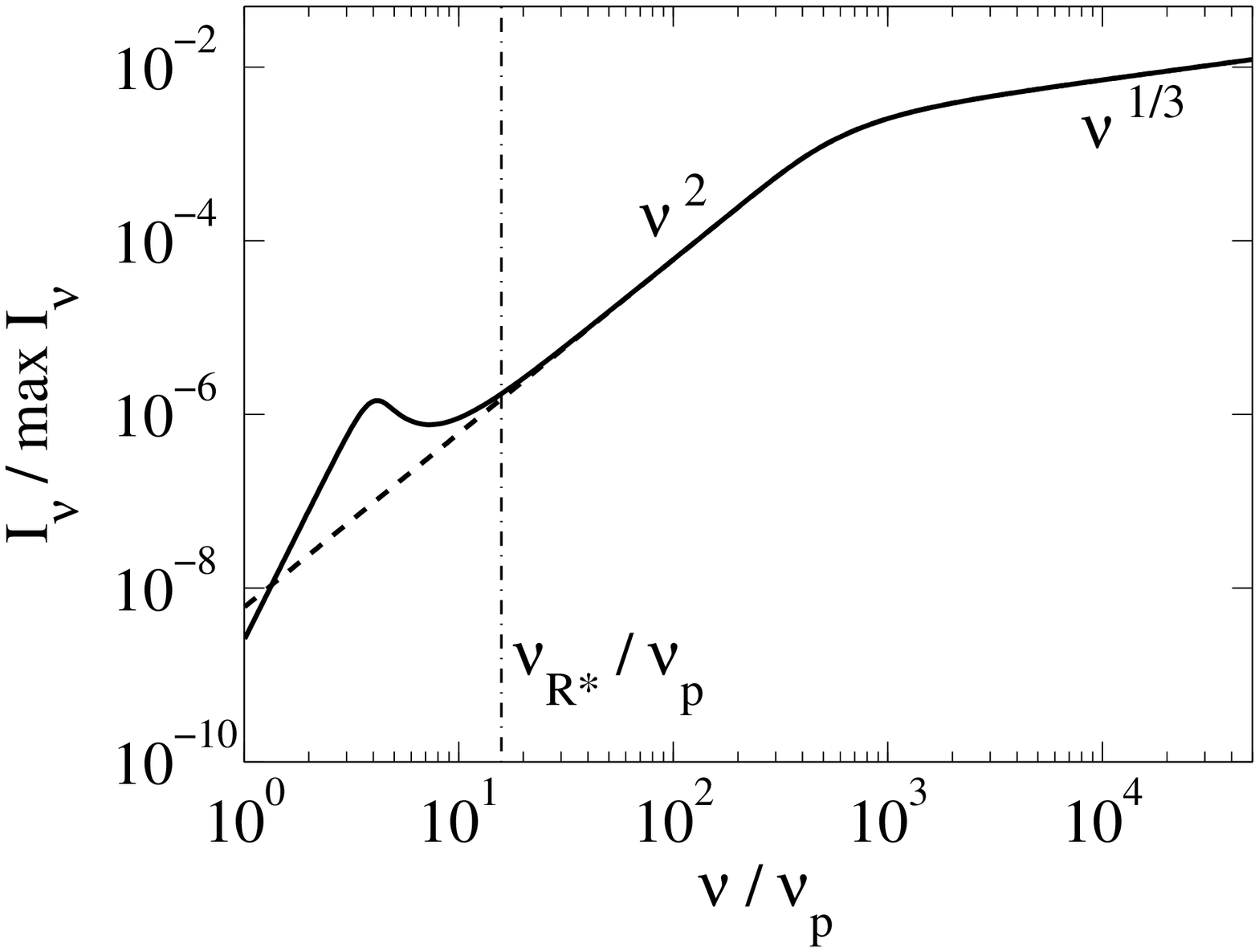}{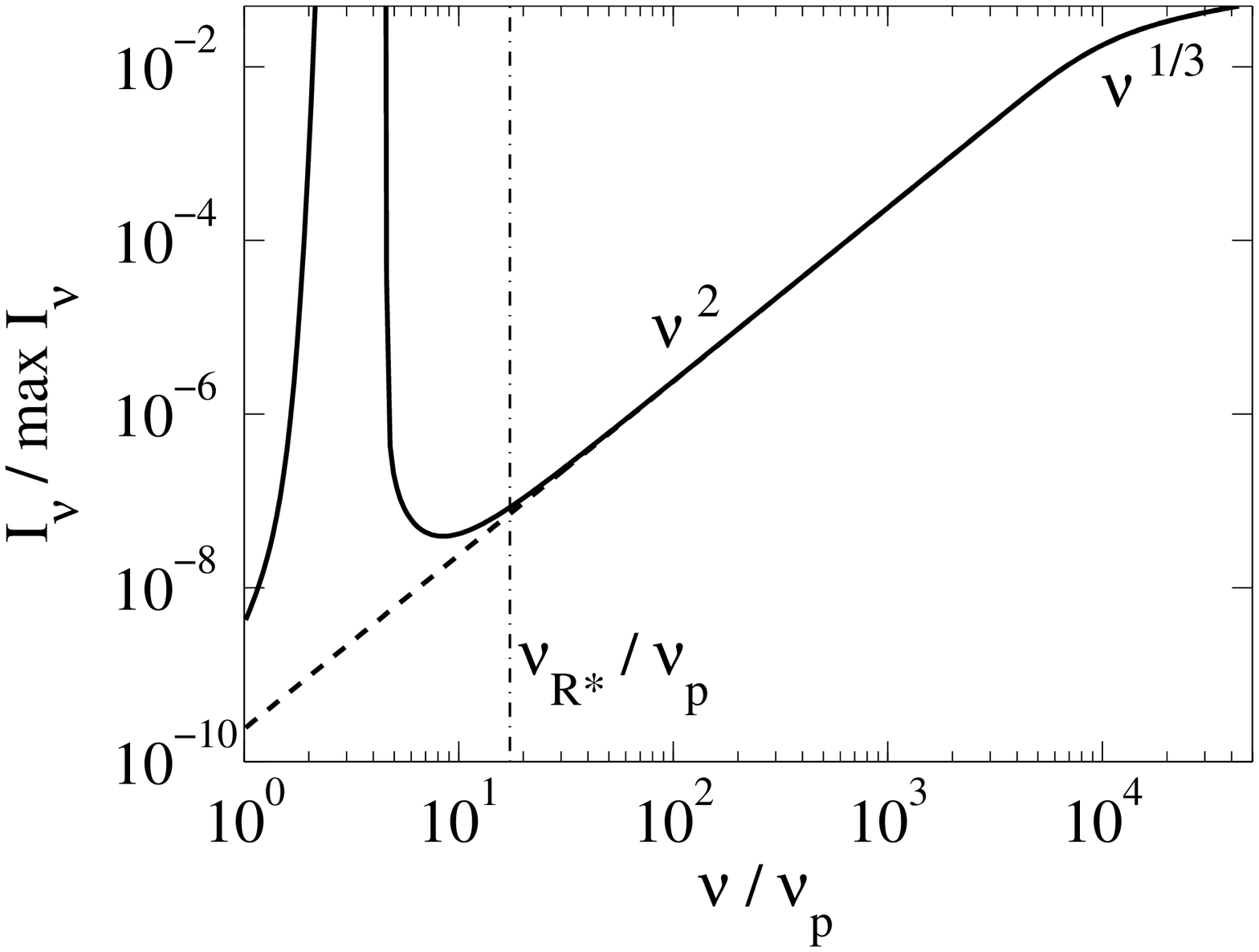}
\figcaption{Emitted intensity of circularly polarized synchrotron radiation
in plasma. \emph{Dashed lines} represent the ``vacuum'' values of $I_{\nu}$, 
and \emph{solid lines} represent these quantities when plasma effects are 
considered.
The vertical \emph{dash-dotted lines} at $\nu_{R^{\scriptstyle{*}}}$ separate 
the region where plasma effects are negligible ($\nu > \nu_{R^{\scriptstyle{*}}}$) 
from the region where these effects are important 
($\nu \lesssim \nu_{R^{\scriptstyle{*}}}$).
In order to demonstrate the plasma collective effects given different plasma
conditions, two parameter sets were used for the calculations presented here.
\newline\emph{Left}~: Typical plasma conditions behind a forward shock propagating into a 
uniform density ISM ($\nu_p\simeq2\times10^5$~Hz in the observer's frame, 
$\gamma_{e\, min} = 4\times10^4$,
$n_e^\prime = 10^3$ and $\gamma_{cool} / \gamma_{e\, min} \simeq 2\times10^3$).
Although there is no negative self-absorption, a $\sim1.5$ order of magnitude
increase in the emitted intensity (compared to the emission in vacuum) is apparent 
at $\nu \lesssim \nu_{R^{\scriptstyle{*}}}\simeq10^3 \nu_p $ 
(see \S\ref{sec:DistributionFunction}).  
\newline\emph{Right}~: Typical plasma conditions behind a forward shock propagating into
a wind ($\nu_p\simeq10^7$~Hz as measured in the observer's frame, $\gamma_{e\, min} = 
10^4$, $n_e^\prime = 10^7$ and $\gamma_{cool} / \gamma_{e\, min} \simeq 10$).
The divergence of the emission at $\nu \lesssim \nu_{R^{\scriptstyle{*}}}\simeq
10\nu_p$ is the signature of the maser effect.
\label{fig:results}} 
\end{figure}

Using a shock injected electron  distribution function consisting of two pure 
power laws with typical afterglow parameters, we verified that the cooling 
scheme 
suggested in \S\ref{sec:DistributionFunction} results in a maser effect at 
frequencies $\nu \lesssim \nu_{R^{\scriptstyle{*}}}$. More realistic distribution
functions are expected to have a smooth transition region between the two 
power laws. As an example, we considered the following shock injected electron 
distribution function (following Gruzinov \& Waxman 1999):
\begin{equation}
n_e^\prime(z) \propto z^2 (1 + a z^{-(p+2)})^{-1}\;,
\end{equation}
where $z=\gamma/\gamma_{e\,min}$. 
The constant $a$ and the overall proportionality factor are chosen such that the 
distribution is adequately normalized. 
This distribution function, 
behaves asymptotically as $\gamma^2$ at low energies and as $\gamma^{-p}$ 
at high energies, with $p = 2.4$. 
We applied the cooling scheme described in \S\ref{sec:DistributionFunction} 
to two sets of plasma parameters~: 
the one, with 
$\nu_p/\nu_B=250, \gamma_{e\,min}=4\times10^4, n_e^\prime=10^3$~cm$^{-3}$ and 
$\gamma_{cool}/\gamma_{e\,min} \simeq 2\times10^3$, 
and the other, with 
$\nu_p/\nu_B=250, \gamma_{e\,min}=10^4, n_e^\prime=10^7$~cm$^{-3}$ and
$\gamma_{cool} \simeq 10\gamma_{e\,min}$. 
Here $\gamma_{cool}=6\pi m_e c/\sigma_T B^2 t'_{dyn}$ is the maximal electron 
Lorentz factor allowing significant synchrotron losses within the dynamic time scale 
of the system, $t'_{dyn}$. 
The first parameter set corresponds approximately to the conditions prevailing 
in the ISM heated by the forward shock, when the fireball expands into a 
uniform-density ISM, while the other corresponds approximately to the physical 
conditions in the plasma accelerated by the forward shock, when the fireball 
expands into a wind. The results of the numerical calculations are displayed in 
Figure \ref{fig:results}. 
When the first parameter set is used (Figure \ref{fig:results} \emph{Left}), 
there is no negative reabsorption, since the cooling is not efficient enough. 
An increase of $\sim1.5$ orders of magnitude in the emitted intensity, compared to 
the emission in ``vacuum'', is apparent, as explained in 
\S\ref{sec:DistributionFunction}. On the other hand, the second parameter set,
which corresponds to a much lower value of $\gamma_{cool}/\gamma_{e\,min}$,
(Figure \ref{fig:results} \emph{Right}) clearly shows the maser effect at frequencies 
$\nu \lesssim \nu_{R^{\scriptstyle{*}}}$, reflecting the negative self absorption 
coefficient due to the efficient synchrotron cooling.

\section{Discussion}\label{sec:Discussion}

We have derived a dispersion relation for transversal electromagnetic waves in 
a weakly magnetized relativistic plasma, assuming the electron and proton 
distribution functions are isotropic. The frequency range at which we were 
interested was $\nu_p<\nu\ll\gamma_e\nu_p$. 
Treating the external weak magnetic field as a perturbation, we have shown 
that at this frequency range the normal modes are circularly polarized, having 
refractive indices which can be approximated by the familiar expression 
$n^2_{1,2}\simeq 1 - (\nu/\nu_p)^2$, where 
$\nu_p = \left[(\nu_{pe}^{NR})^2/\gamma_e + (\nu_{pp}^{NR})^2/\gamma_p 
\right]^{1/2}$. 
We have shown that this result is valid whether the electron distribution is 
assumed to be a $\delta-$function or a power-law. We have also calculated the 
difference between the two refractive indices [see eq. (\ref{eq:Delta_n_revised})], 
obtaining 
$|\Delta n| \simeq (\nu_p^2 \nu_B/\nu^3) \cos\phi\ \ln \left(\nu_p^2/4\nu^2\right)$, 
for radiation propagating at angle $\phi<|\pi/2 - 0.1 \nu_B \nu/\nu_p^2|$ 
with respect to the external magnetic field, and $|\Delta n| \simeq \nu_B^2/2\nu^2$ 
for radiation propagating at larger angles (i.e., almost perpendicular to the 
field direction). 

We next came to consider the effects of the relativistic plasma on synchrotron emission. 
We have derived an estimate of the frequency at which the plasma effect on the emission 
becomes significant~:
\begin{equation}
\nu_{R^{\scriptstyle{*}}} \simeq \nu_p \min \left\{\gamma, 
\sqrt{\frac{\nu_p}{\nu_B}}\right\}. 
\end{equation}
This result generalizes the familiar \emph{Razin} frequency $\nu_R=\gamma_e\nu_p$ 
to a regime where $\nu_p/\nu_B\ll\gamma_e^2$, which is the relevant regime for typical 
plasma parameters at GRB afterglows; hence we expect strong effect of the plasma on the 
synchrotron emission at $\nu_{R^{\scriptstyle{*}}} = \nu_p \sqrt{\nu_p/\nu_B}$. 
Consequently we found that a necessary condition for the collective plasma effects to 
be observable is $\nu_p/\nu_B \gg 1$. This ratio was shown to depend only on the 
ratio of $\xi_e$ to $\xi_B$, and on the width of the distribution function 
[eq. (\ref{eq:nu_R*})]. Hence $\xi_e/\xi_B\gg 1$ is a necessary condition 
for the effects to be observable.

We applied the above results to  plasma parameters 
($\gamma_e \simeq 10^3-10^4, \nu_p/\nu_B \simeq 10^2-10^3$) typical to GRB 
afterglows, and found that for $\xi_B \simeq 10^{-6}$ and $\xi_e$ close to 
equipartition, the plasma had a decisive effect on the synchrotron emission at 
radio frequencies, during the transition of the fireball dynamics to 
self-similarity. Two expansion scenarios were considered~: 
(i) a blast wave into a uniform-density ISM with $n_e \simeq 1$~cm$^{-3}$, 
where we expect a plasma collective effect on synchrotron emission from the gas 
heated by the forward shock at $\sim3.5$~MHz; 
(ii) expansion into a wind, where the number density decreases as $r^{-2}$, 
and at the radius at which the dynamics becomes self-similar is 
$\sim 10^4$~cm$^{-3}$. In this case we predict a plasma effect on synchrotron 
emission from the wind accelerated by the forward shock at $\sim0.2$~GHz 
[see eqs. (\ref{eq:UniformISM_ForwardShock})-(\ref{eq:Wind_ReverseShock})].

Note that plasma collective effects on synchrotron emission below 
$\nu_{R^{\scriptstyle{*}}}$ are expected irrespective of the details of 
the electron distribution function. This is in contrast to negative reabsorption 
and the maser effect, which require a region in the electron distribution function 
which rises faster than $E^2$. The energy distribution of electrons 
produced by collisionless shock acceleration is not well known, in particular at low 
energy. However, we have shown that for plausible assumptions on the electron 
distribution function at low energy, synchrotron cooling of the electrons may lead 
to negative reabsorption. This, however, depends on the details of the distribution 
function near its peak. We have demonstrated negative reabsorption for a
distribution which transforms smoothly (at $\gamma_{e\,min}$) from a 
$E^2$ power-law at low energies to a $E^{-p} \;(p = 2.4)$ power-law at high 
energies, for a certain range of parameters 
(Figure \ref{fig:results} \emph{Right}). 

One should notice, however, that even if the distribution function permits negative 
reabsorption, coherent emission from plasma accelerated by a reverse shock may be 
completely obscured by the high optical depth of the ambient medium heated by the 
forward shock. Thus, for example, when the fireball expands into a wind, certain 
electron distribution functions may lead to coherent emission from the reverse shock 
at 1.2~GHz [eq. (\ref{eq:Wind_ReverseShock})]. Nevertheless, the maser due to the 
reverse shock is completely suppressed by the $\sim10^3$ optical depth of the 
forward shock at the GHz frequency range. A similar situation occurs also when the 
fireball expands into a uniform-density ISM. Note, however, that if turbulent 
processes disrupt the shock fronts, so that ``bulges'' of accelerated ejecta lie 
in front of the forward shell along the line of sight, the plasma effects occurring 
in the reverse shock may become observable. 

Our calculations of coherent emission assumed a homogeneous distribution function in 
the emitting shell. However, we have numerically shown that even if inhomogeneity due 
to different cooling times is taken into account, coherent emission would still be 
observable~: the regions close to the shock front, where the electrons have little 
time to cool, contribute only a positive optical depth of order of few, hence do 
not obscure completely the maser emission from regions further away from the shock 
front.

We have shown that for typical plasma parameters the normal waves propagating in the 
plasma are circularly polarized at frequencies of order of $\nu_{R^{\scriptstyle{*}}}$
[see the discussion following eq. (\ref{eq:IsotropyConsistencyCondition}) and 
\S\ref{sec:SynchrotronPolarization}]. 
Since the difference between the refractive indices of the transversal normal modes is 
smaller by a factor of $\sim \nu_B/\nu \ll 1$ than $1-n_{1,2}(\nu)$, and it is 
the latter which enter the equations of reabsorption and determine the amplification of 
radiation in both polarizations, the resulting optical depths corresponding to radiation 
in the two polarizations differ by a factor of a few, at most, if the magnetic field is 
assumed to be constant throughout the emitting shell. This factor is expected to reduce, 
due to saturation of emitted radiation. If, however,  the direction of the magnetic field 
is assumed to fluctuate randomly on length-scales smaller than the shell's width, the 
difference between the two refractive indices also changes signs randomly, and the 
difference in the amplifications of the two polarization is averaged out. 
Consequently, we do not expect the observed radiation to have a high degree of circular 
polarization at relevant frequencies.  

A detection of a strong effect in the radio-band at early stages of the afterglow will 
place a new independent constraint on the value of $\xi_B$, as it will prove that it 
must be much smaller than 1. 
Since there is a two orders of magnitude difference between our estimates of 
$\nu_{R^{\scriptstyle{*}}}$ in the forward shock in the two expansion scenarios, 
a flash in radio wave-band during early afterglow stages seems more probable when the 
fireball expands into a wind. In this respect, plasma collective effects may serve as 
an important clue to the GRB environment and progenitor type.

\acknowledgements
EW is the incumbent of the Beracha foundation career development chair.

\appendix
\section{Dispersion relations of transversal electromagnetic waves 
in a relativistic plasma}
\label{sec:App_A}

Since the rate of binary collisions is much smaller than the frequencies of interest 
for us, the plasma may be adequately described by a collisionless Vlasov equation~:
\begin{equation}\label{eq:Vlasov}
\frac{\partial f_{\alpha}}{\partial t} + \frac{\mathbf{p}}{\gamma m_{\alpha}} 
{\mathbf\cdot \nabla} f_{\alpha} + q_{\alpha} \left( {\mathbf{E}} + 
\frac{{\mathbf p \times B}}{\gamma m_{\alpha} c} \right) 
{\mathbf \cdot} \frac{\partial f_{\alpha}}{\partial \mathbf{p}} = 0\;,
\end{equation}
where the one-particle distribution function $f_{\alpha} ({\mathbf x}_{\alpha}, 
{\mathbf p}_{\alpha}, t) $ is defined as the fractional density of particles of a 
single type in phase space, and $\alpha$ denotes a single species of particles, 
i.e. electrons or protons.

\subsection{Transversal EM waves in a field-free plasma}\label{sec:FieldFreePlasma}

Let $f_{\alpha 0}$ be an equilibrium distribution function, and let $f_{\alpha 1} 
({\mathbf x},{\mathbf p},t)$ be a small perturbation to it, so that $f_{\alpha} = 
f_{\alpha 0}+f_{\alpha 1}$. It is assumed that the equilibrium distribution 
function carries no net charge or current distributions. Since we assume that 
$|f_{\alpha 1}| \ll f_{\alpha 0}$, equation (\ref{eq:Vlasov}) can be linearized, 
and we obtain the (Fourier-transformed) perturbation to the equilibrium distribution 
function 
\begin{equation}\label{eq:f_alpha1}
f_{\alpha 1} = q_{\alpha} \frac{{\mathbf E \cdot}(\partial f_{\alpha 0}/ 
\partial{\mathbf p})}{i \left(\omega-{\mathbf k\cdot p}/\gamma m_{\alpha}\right)}\;,
\end{equation}
where we have used the assumption that $f_{\alpha 0}$ is isotropic.  
Here $\omega=2\pi\nu$. 
The transversal electromagnetic fields satisfy Maxwell's equations
\begin{eqnarray}\label{eq:Maxwell}
{\mathbf \nabla \times E} &=& -\frac{1}{c} 
\frac{\partial {\mathbf B}}{\partial t}\nonumber\\
{\mathbf \nabla \times B} &=& \frac{1}{c} 
\frac{\partial {\mathbf E}}{\partial t} + 
\frac{4 \pi}{c} {\mathbf j}\;,
\end{eqnarray}
where the net current is given by ${\mathbf j} = \sum_{\alpha} \bar{n}_{\alpha} 
q_{\alpha} \int ({\mathbf p}/\gamma m_{\alpha}) f_{\alpha 1} d^3p$, the average 
number density of the particles of the species $\alpha$ is $\bar n_{\alpha}$, 
and $\gamma=\sqrt{p^2+m_{\alpha}^2 c^2}/m_{\alpha} c$ is the particle's Lorentz 
factor. We choose a system of axes such that the radiation propagates with a wave 
vector ${\mathbf k}=(0,0,k)$ along the $z$ axis, whence the transversal electric 
field is confined to the $xy$ plane~: ${\mathbf E}=(E_{\perp 1},E_{\perp 2},0)$, 
and the particle momentum vector is ${\mathbf p}=(p_{\perp 1},p_{\perp 2},
p_{\parallel})$. Equation (\ref{eq:Maxwell}) leads to a dispersion relation for 
the transversal waves~:
\begin{equation}\label{eq:Dispersion2x2}
k^2 c^2 / \omega^2 E_i = \epsilon_{ij} E_j \quad\mbox{with}\quad i,j = 1,2
\end{equation}
where $\epsilon_{ij}$ is the dielectric tensor~:
\begin{mathletters}
\begin{eqnarray}\label{eq:epsilon11}
\epsilon_{11} = \epsilon_{22} &=& 1 + \displaystyle\sum_{\alpha} 
\frac{(\omega_{p \alpha}^{NR})^2}{\omega} \int d^3p 
\frac{1}{(\omega - k v_{\parallel})} 
\frac{d f_{\alpha 0}}{d p} \frac{p_{\perp 1}^2}{p} 
\frac{m_{\alpha} c}{\sqrt{p^2 + m_{\alpha}^2 c^2}} \\
\epsilon_{12} = -\epsilon_{21} &=& 0\;.
\end{eqnarray}
\end{mathletters}
Here $\omega_{p \alpha}^{NR}=(4\pi\bar n_{\alpha} q_{\alpha}^2/m_{\alpha})^{1/2}$ 
is the non-relativistic plasma frequency of the species $\alpha$. 
The refractive index $n(\omega)= kc/\omega$ for the transversal waves is the 
square root of the doubly-degenerate eigenvalue of equation 
(\ref{eq:Dispersion2x2}).

If the electron and proton distribution functions can be approximated by 
$\delta-$functions $f_{\alpha 0}=(1/4\pi p_{\alpha 0}^2) 
\delta(p-p_{\alpha 0})$, equation (\ref{eq:epsilon11}) becomes
\begin{eqnarray}\label{eq:epsilon11_DeltaFunction_FieldFree}
\epsilon_{11} = \epsilon_{22} &=& 1 - \displaystyle\sum_{\alpha} 
\frac{(\omega_{p \alpha}^{NR})^2}{2 \gamma_{\alpha 0}\omega^2} 
x_{\alpha} \nonumber\\
&\times& \Biggl\{\left( x_{\alpha}^2 \left( 1-\frac{1}
{\gamma_{\alpha 0}^2}\right)-1\right)\ln \left| 
\frac{x_{\alpha}-1}{x_{\alpha}+1} \right| + 2 x_{\alpha} 
\left(1 - \frac{1}{\gamma_{\alpha 0}^2}\right)\Biggr\}\;,
\end{eqnarray}
with $x_{\alpha} = \omega / k v_{\alpha 0}$, and 
$v_{\alpha 0} = p_{\alpha 0} / \gamma_{\alpha 0} m_{\alpha}$. 
It can be shown that in the frequency range of interest, and for relevant values 
of $\gamma_{e,p}$, the second term in the curly brackets of equation 
(\ref{eq:epsilon11_DeltaFunction_FieldFree}) dominates the first term for both 
the electrons and the protons. This may be used to approximate the refractive 
index by~: 
\begin{mathletters}
\begin{eqnarray}\label{eq:n_approx}
n_{app}^{\phantom{a}2}(\omega) &=& 1-\left( \frac{\omega_p}{\omega}\right)^2\;,\\ 
\omega_p &=& \left[ \frac{4 \pi \bar{n}_e e^2}{\gamma_{e\,0} m_e} 
+ \frac{4 \pi \bar{n}_e e^2}{\gamma_{p\,0} m_p} \right]^{1/2} = 
\left[ \frac{(\omega_{pe}^{\phantom{p}NR})^2}{\gamma_{e\,0}} 
+ \frac{(\omega_{pp}^{\phantom{p}NR})^2}{\gamma_{p\,0}} \right]^{1/2} \;.
\end{eqnarray}
\end{mathletters}

Both observations and theory imply that in afterglow plasmas the electrons are not 
distributed mono-energetically, but rather have a distribution function which extends 
to high energies as a power-law, with a spectral index $p \gtrsim 2$. 
However, we have shown numerically that in the frequency range of interest for us, 
where $|1-(kc/\omega)| \gg 1/\gamma_e^2$, and for relevant values of electron and 
proton Lorentz factors, $|1-n_{PL}|$ differs from $|1-n_{\delta}|$ by less than 
one part in $10^4$, where $n_{\delta}$ and $n_{PL}$ are the refractive 
indices for the mono-energetic and power-law distribution functions, respectively. 
Consequently, result (\ref{eq:n_approx}) still holds approximately for a power-law 
distribution. Hereafter we shall  use $\delta-$function distributions to evaluate 
the deviation of the refractive index from 1.

\subsection{Transversal EM waves in a weakly magnetized plasma}
\label{sec:WeaklyMagnetizedPlasma}

Let $f_{\alpha 0}$ be an isotropic equilibrium distribution function, which 
describes a plasma in an external uniform magnetic field ${\mathbf B}_0$. 
We introduce a small perturbation to $f_{\alpha 0}$. 
Since we are interested in frequencies $\omega \gg \omega_B$, we assume that 
the external magnetic field is small, so that the perturbation  is made up of two 
contributions, one (which we denote by $f_{\alpha 1}$) which is independent of 
the magnetic field, and the other (denoted by $f_{\alpha 2}$) which is linear 
in the magnetic field; thus $f_{\alpha 0} \gg |f_{\alpha 1}| \gg |f_{\alpha 2}|$. 
We now linearize the Vlasov equation [see eq. (\ref{eq:Vlasov})], and neglect 
$(\partial f_{\alpha 2} / \partial {\mathbf p})$ with respect to 
$(\partial f_{\alpha 1} / \partial {\mathbf p})$. 
Substituting equation (\ref{eq:f_alpha1}) for $f_{\alpha 1}$, 
we obtain (after some algebra) 
\begin{equation}\label{eq:f_alpha2}
f_{\alpha 2} = - \frac{q_{\alpha}^2}{\gamma m_{\alpha} c}  
\frac{(d f_{\alpha 0} / d p)}{p} ({\mathbf p \times B}_0) {\mathbf \cdot} 
\left[ \frac{{\mathbf E}}{\left(\omega-{\mathbf k \cdot p}/\gamma m_{\alpha} 
\right)^2} + \frac{({\mathbf E \cdot p}){\mathbf k}}{\gamma m_{\alpha} 
\left(\omega - {\mathbf k\cdot p}/\gamma m_{\alpha}\right)^3 }\right]\;. 
\end{equation}
Hence $f_{\alpha 2}$ is indeed linear in the external magnetic field, 
consistent with our assumption.

As already mentioned, we are interested in frequencies $\omega\gg\omega_B$. 
For these frequencies the effect of the magnetized plasma on the propagating 
radiation is small, and the electric field can be assumed to be approximately 
transverse, i.e. ${\mathbf E}\perp{\mathbf k}$. Our previous choice of axes is 
convenient for the analysis; without any loss of generality, we choose 
${\mathbf B}_0 = (0,B_0 \sin\phi, B_0 \cos \phi)$. 
Following the steps outlined in equations (\ref{eq:Maxwell}) - 
(\ref{eq:epsilon11}), we obtain the dielectric tensor for quasi-transversal 
EM waves in a weakly magnetized plasma~:
\begin{equation}\label{eq:epsilon12}
\epsilon_{12} = -\epsilon_{21} = i \displaystyle\sum_{\alpha} 
\frac{(\omega_{p \alpha}^{NR})^2}{\omega} {\omega}_{B \alpha}^{NR}  
\cos\phi \int d^3p  \frac{1}{( \omega - k v_{\parallel} )^2} 
\frac{d f_{\alpha 0}}{d p}  \frac{p_{\perp 1}^2}{p} 
\frac{m_{\alpha}^2 c^2}{p^2 + m_{\alpha}^2 c^2} \;,
\end{equation}
with ${\omega}_{B \alpha}^{NR} \equiv |q_{\alpha}|B_0/m_{\alpha}c$ 
the non-relativistic gyration frequency. The expression for 
$\epsilon_{11} = \epsilon_{22}$ are given in equation (\ref{eq:epsilon11}). 
For mono-energetic electron and proton distributions, equation 
(\ref{eq:epsilon12}) becomes
\begin{eqnarray}\label{eq:epsilon12_DeltaFunction}
\epsilon_{12} = -\epsilon_{21} &=& i \displaystyle\sum_{\alpha}  
\frac{(\omega_{p \alpha}^{NR})^2 {\omega}_{B \alpha}^{NR}}{\omega^3} 
\frac{\cos\phi}{2 \gamma_{\alpha 0}^2} x_{\alpha}^2 
\Biggl\{ \left( 1 - \frac{1}{\gamma_{\alpha 0}^2} \right) x_{\alpha} 
\ln \left| \frac{x_{\alpha}-1}{x_{\alpha}+1} \right| \nonumber\\
&& -  \frac{2}{\gamma_{\alpha 0}^2} \frac{1}{x_{\alpha}^2-1} 
- 2  - \frac{10}{\gamma_{\alpha 0}^2} \Biggr\} \;,
\end{eqnarray}
where $x_{\alpha}$ has the same definition as in equation 
(\ref{eq:epsilon11_DeltaFunction_FieldFree}).

% -----------------------------------------------------------------------------
% ----------------------------    BIBLIOGRAPHY     ----------------------------
% -----------------------------------------------------------------------------

% -------------------------   END OF BIBLIOGRAPHY     -------------------------

\clearpage

\end{document}